\documentclass{aa}
\usepackage{graphics}
\newcommand{\ksM}{\mbox{$\mathrm{km\,s^{-1}\,Mpc^{-1}}$}}
\begin{document}
\thesaurus{03(11.05.2; 11.19.3; 11.12.2; 13.21.2; 12.03.2)}
\title{Far-UV and deep surveys: bursting dwarfs versus normal galaxies}
\author{Michel Fioc\inst{1,2} \and Brigitte Rocca-Volmerange\inst{1,3}}
\institute{Institut d'Astrophysique de Paris, CNRS,
98 bis Bd Arago, F-75014 Paris, France 
\and NASA/Goddard Space Flight Center, code 685, Greenbelt, MD 20771, USA
\and Institut d'Astrophysique Spatiale,
B\^at. 121, Universit\'e Paris XI, F-91405 Orsay, France}
\offprints{Michel Fioc}
\mail{fioc@iap.fr}
\date{Received / Accepted}
\authorrunning{M. Fioc \& B. Rocca-Volmerange}
\titlerunning{Bursting dwarf galaxies}
\maketitle
\begin{abstract}
Galaxy counts from bright ultraviolet (UV) and deep op\-ti\-cal spectroscopic surveys
have revealed an unexpectedly large number of very 
blue galaxies (vBG). The colors and luminosities of these objects indicate that they 
are dwarf galaxies undergoing bursts of star formation. 
We use a galaxy evolution model (\textsc{p\'egase}, Fioc \& Rocca-Volmerange 1997, hereafter FRV) 
to describe this
population as galaxies undergoing cyclical bursts of star formation, thereby determining
the luminosity function (LF) of these galaxies. 

When these bursting galaxies are added to normally evolving
populations, the combination reproduces the UV number counts,
color distributions and deep optical redshift distributions fairly 
well. Optical (including the Hubble Deep Field) and near-infrared number counts are
fitted assuming an open or a flat, $\Lambda$-dominated, Universe. The high 
amplitude of the angular correlation function of very blue galaxies discovered 
by Landy et al. (1996) is also recovered in this modelling.

The number of bursting galaxies is only a small fraction of the
total number of galaxies at optical and near-infrared wavelengths, 
even at faintest magnitudes. In our evolution mo\-del\-ling,
normal galaxies explain most of the blue excess in a low-$\Omega$ 
Universe. The problem of the blue excess
remains in a flat Universe without a cosmological constant.
\end{abstract}
\section{Introduction}
The apparent excess of the number of galaxies at faint magnitudes in the blue, relative to predictions of 
non-evolving models, is a long-standing problem of cosmology (Koo \& Kron 1992).
Pure luminosity evolution (PLE) scenarios fitting the colors of nearby galaxies 
reproduce the optical and near-infra\-red observations for an open Universe
(Guiderdoni \& Rocca-Volmerange 1990; Pozetti et al. 1996; 
Fioc 1997) 
or for a flat, $\Lambda$-dominated, one (Fukugita et al.~1990; Fioc 1997).
To solve the problem of the blue 
excess in a flat, $\Lambda=0$, Universe, various solutions were proposed 
as a strong number density 
evolution of galaxies via merging (Rocca-Volmerange \& Guiderdoni 1990; Broadhurst et al. 1992), a very steep luminosity function (Driver et 
al.~1994)
or a population of fading dwarf galaxies (Babul \& Rees 1992; Babul \& Ferguson
1996; Campos 1997).

Determining the distance of these galaxies is crucial to understand their 
nature: they should be relatively high-redshift, 
intrinsically bright galaxies according to PLE scenarios, and low-redshift, 
intrinsically faint galaxies in other cases.
Despite the very faint magnitudes now reached in the optical by the Hubble 
Space Telescope (Hubble Deep 
Field (HDF), Williams et al. 1996)
or in the near-infrared (e.g. Moustakas et al. 1997), and the advent of 
complete, deep spectroscopic surveys 
(Lilly et al. 1995; Cowie et al. 1996), 
the question of the blue excess is still open (Ellis 1997).

Far-UV studies can throw new light on this problem. 
By analyzing bright data at 2000~\AA\ from the balloon experiment 
FOCA\-2000,
Armand \& Milliard (1994) showed that the galaxy number-magnitude 
counts predicted with classical  optical luminosity functions and 
typical UV-optical colors as a function of morphology are deficient by a 
factor 2 relatively to the observations.
Moreover, the color distribution indicates that this excess is 
due to very blue galaxies ($UV-B\sim -2$). 
A question immediately arises: is the UV excess related to the blue excess?
Gronwall \& Koo (1995) introduced {\em non-evolving populations} of 
faint {\em very} blue galaxies, contributing significantly to faint counts
in order to find the luminosity function (LF) that best fitted the observational constraints. 
A similar population was proposed by Pozzetti et al. (1996),
but with a much smaller contribution, in order to fit the $B-R$ color distributions.
However, very blue colors require
that {\em individual} galaxies are bursting and are therefore rapidly evolving. 
By modelling the spectral evolution of these galaxies taking in 
consideration post-burst phases,
Bouwens \& Silk (1996) concluded that the LF adopted by Gronwall \& Koo (1995)
leads to a strong excess of nearby galaxies in
the redshift distribution and that vBG may not be the main explanation of
the blue excess. Such a conclusion was also drawn by Koo (1990).

In this paper we combine the findings of recent deep surveys with the new UV data 
(Sect.~2) in order to develop a model of vBG and to determine their luminosity function (Sect.~3).
We then discuss 
their contribution to faint galaxy counts (Sects.~4 and~5), depending
on the cosmology. 
\section{Observational evidence for very blue galaxies}
In contrast with the so-called `normal' galaxies of the Hubble sequence,
which are supposed to form at high redshift with definite star formation timescales as in FRV,
bursting galaxies evolve rapidly without clear timescales.
In the red post-burst phases they 
might be indistinguishable from normal
slowly evolving galaxies. However they can be identified during their bluest,
bursting phase, allowing determining their evolution and number density.

\begin{figure}
\resizebox{\hsize}{!}{\rotatebox{-90}{\includegraphics{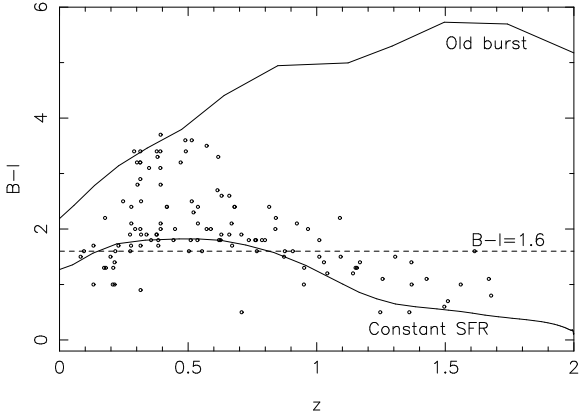}}}
\caption{$B-I$ versus $z$ for galaxies from Cowie et al. (1996) sample.
The thick lines define the envelope of normal galaxies. The upper one holds 
for a 13 Gyr-old initial burst without subsequent star formation and the 
lower one for a 10 Gyr-old galaxy forming stars at a constant rate. The 
dashed line separates galaxies at $B-I=1.6$. A significant fraction 
of galaxies are observed outside the envelope at $z\sim0.2$, with $B-I<1.6$.}
\label{cowie}
\end{figure}

The existence of a population of galaxies much bluer than normal and classified as starbursts
has been recently noted
at optical wavelengths by Heyl et al. (1997, hereafter HCEB). At fainter magnitudes ($B=22.5-24$), 
the Cowie et al. (1996) deep survey has revealed two populations of blue ($B-I<1.6$) galaxies 
(Figs.~\ref{cowie} and \ref{nz}). 
Normal star forming galaxies, as predicted by standard PLE models, are observed at high redshift 
($z>0.7$), but another clearly distinct population of blue galaxies is identified 
at $0<z<0.3$. Some of these galaxies are very blue. 
This second population was previously observed in the brighter survey 
($b_{\mathrm{j}}<22.5$) of Colless et al. (1993) and was thought to be the cause
of the blue excess, in the absence of high-redshift galaxies. More recently,
Roche et al. (1997) observed two kinds of blue galaxies at $I<24$: a 
population with a number density in agreement with PLE models, and a subset of vBG
with small angular sizes.

The best constraint on vBG comes from 
the far-UV (2000 \AA) bright counts observed with the balloon experi\-ment 
FOCA\-2000 (Armand \& Milliard 1994).
By using a standard optical LF, the authors obtain a strong deficit of predicted 
galaxies 
in UV counts across the magnitude range ($UV=14-18$) and argue in favor 
of a LF biased toward
later-type galaxies. 
\begin{figure}
\resizebox{\hsize}{!}{\rotatebox{-90}{\includegraphics{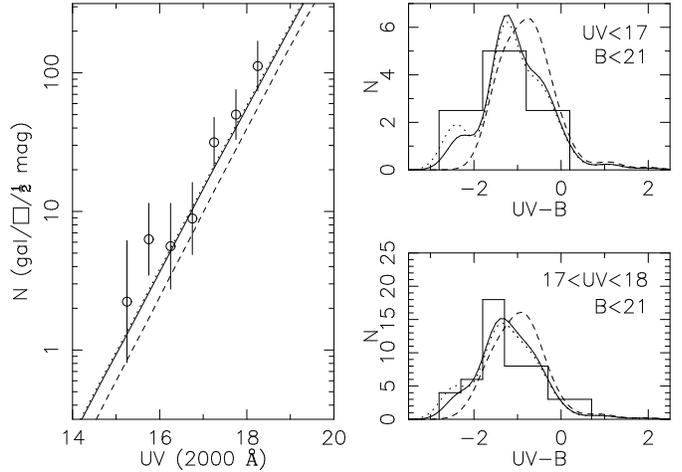}}}
\caption{Number counts and color distributions predicted with Heyl et al. 
(1997) LF, without inclusion of bursting galaxies (dashes), and with them
(solid: Rana \& Basu (1992) IMF; dots: Salpeter (1955) IMF), 
compared to the observations
of Armand \& Milliard (1994) (circles and histograms). 
Color distributions are 
normalized to the area of the histograms.} 
\label{UV}
\end{figure}
In FRV, we used the LF of Marzke et al. (1994), while here
we prefer to adopt the LF of HCEB as it contains a higher
fraction of star-forming galaxies, favoring a better fit to the 
data.
Even with this new LF, the star formation scenarios proposed in FRV predict the same factor
of 2 deficit in UV counts (Fig.~\ref{UV}, dashed line) as that found by Armand \& Milliard (1994).
This is despite the rather
high normalization of the LF to the bright multispectral galaxy counts of Gardner et al. 
(1996) and Bertin \& Dennefeld (1996) (see the discussion in FRV).
Moreover, the predicted $UV-B$ color distributions show
a clear lack of blue galaxies, notably of those with $UV-B<-1.5$ 
(Fig.~\ref{UV}, dashed lines).
A 10 Gyr-old galaxy forming stars at a constant rate, would 
only have $UV-B\sim-1.2$. 
Although lowering the metallicity of the models may lead to bluer colors (Fioc \& Rocca-Volmerange
1998, in preparation), they will 
still be too red to explain the data.
A population of bursting galaxies is clearly needed to explain UV counts 
and the Cowie et al. (1996) data.  
\section{Modelling very blue galaxies}
\subsection{Star formation scenario}
Galaxies can have very blue colors either if they are very young or if they are 
undergoing enhanced star formation.
Two kinds of models have been advanced by
Bouwens \& Silk (1996) to maintain a population of vBG over a wide range of redshifts. In the first one, 
new blue galaxies form continuously, leaving red fading remnants, whereas in the second,
star formation occurs recurrently. 
The second scenario has both observational and theoretical support and is
adopted in the following. 
Smecker-Hane et al. (1996) concluded to an episodic star formation rate (SFR) from the analysis 
of the stellar populations observed in the Carina dwarf spheroidal galaxy.
Recurrent star formation may also provide an attractive explanation of the
existence of various types of dwarf galaxies, as proposed by Davies \&
Phillips (1988): blue compact dwarfs may correspond to bursting phases
and dwarf irregulars to quiescent ones.
According to the stochastic self propagation star formation theory (Gerola 
et al. 1980), episodic star formation may be a common 
phenomenon (Comins 1984). Such a behavior should be more frequent in dwarf 
galaxies than in giant galaxies, since the probability of propagation of
star formation increases with galaxy mass (Coziol 1996).
The feedback of massive stars on the interstellar medium via 
supernovae and winds may also lead to oscillations of the SFR
(Wiklind 1987; 
Firmani \& Tutukov 1993; Li \& Ikeuchi 1988).
Finally, in the models where the SFR is controlled by the interactions 
with the environment (Lacey et al. 1993), the lower frequency of
interactions of dwarf galaxies may result in an episodic star formation.

For the sake of simplicity, we assume that 
all bursting gala\-xies form stars periodically. In each period, a burst phase with a constant SFR 
$\tau$ and the initial mass function  
(IMF) of Rana \& Basu (1992)
used in FRV for normal galaxies
is followed by a quiescent phase without star formation.
We do not apply any extinction to these galaxies. Because they are
dwarf, we expect
them to have a low metallicity and, therefore, to suffer little extinction
(Heckman et al. 1996).
Moreover, the extinction law (Calzetti et al. 1995), 
the respective distributions of the dust and stars (Gordon et al. 1997)
and the ability of dwarf galaxies to retain
the metals expelled by dying stars (e.g. Dekel \& Silk 1986), as well as the dilution factor of
these ejecta in the interstellar medium, are still under debate.
Neglecting the extinction is clearly a critical assumption, but also a 
conservative one since any
extinction would redden the spectrum and force us to adopt more extreme 
star formation parameters to recover very blue colors. For the same reason,
we do not consider the hypothesis of a low upper mass cut-off of the IMF,
which has been suggested by Doyon et al. (1992), but is controversial
(Lan\c{c}on \& Rocca-Volmerange 1996, Vacca et al. 1995).
In the opposite, an IMF enriched in massive stars, either because
of a flatter slope or because of a high lower mass cut-off, as advocated
for some starbursts (Puxley 1991, Rieke et al. 1993), is more 
attractive since burst phases will be bluer.
Recent studies (Calzetti 1997, Scalo 1997, Massey \& Armandroff 1995) seem 
actually to favor a rather standard IMF, though maybe slightly
flatter at high mass than the one used in FRV.
Such an IMF, e.g. Salpeter (1955), may indeed provide a better fit
to the UV data (Fig.~\ref{UV}, dotted line) than our standard IMF (Rana \& 
Basu 1992, solid line). 
The Salpeter IMF however gives very similar results for other galaxy counts
and even slightly overpredicts the $(F300W-F450W)$ color 
distribution of HDF galaxies in the blue.
For this reason, and also to preserve the consistency with that of normal 
galaxies, we prefer in the following to adopt the Rana \& Basu IMF
for all galaxies.

A good agreement with observational constraints is obtained with 100 Myr-long 
burst phases occurring every 1 Gyr. Surprisingly, similar values were 
adopted by Olson \& Pe\~na (1976) to fit the colors of the Small Magellanic 
Cloud. One should however be aware that these values are very uncertain, 
depending particularly on the metallicity of stars, and are simply mean values
aimed at reproducing the $UV-B$ color distribution.
\subsection{Luminosity function}
Because bursting galaxies evolve very rapidly and have thus a large variety 
of luminosities and colors at every redshift, we distribute them in subtypes
as a function of the time elapsed since the beginning of the last burst.
The LF of each subtype is derived from the SFR function $\Phi(\tau)$ 
that we parameterize,
by analogy with the LF of normal galaxies, as
\[\Phi(\tau)\mathrm{d}\tau=\phi_{\tau}^{\ast}
\exp\left(-\frac{\tau}{\tau^{\ast}}\right)
\left(\frac{\tau}{\tau^{\ast}}\right)^{\alpha_{\tau}}
\frac{\mathrm{d}\tau}{\tau^{\ast}}.\]

The lack of vBG at $z\ga0.4$ in the Cowie et al. (1996) redshift distribution  
(Fig.~\ref{nz})
strongly constrains the LF. This deficit may be interpreted in two 
ways. Either bursting galaxies formed only at low redshifts ($z<0.4$), or the 
lack of vBG may correspond to 
the exponential cut-off of the adopted Schechter function. 
Physical arguments for low redshifts of galaxy formation are weak.
Scenarios invoking a large population of blue dwarf galaxies, as proposed by
Babul \& Rees (1992), generally predict a higher redshift of formation ($z\sim 1$). Adopting the 
second explanation, we get 
$M^{\ast}_{\rm b_j}\sim-17$ (for $H_0=100\,\ksM$) for the overall LF of
bursting galaxies at $z=0$ and can constrain the 
other parameters.
A steep LF extending to very faint magnitudes leads to a large local ($z<0.1$) excess
in the redshift distribution (Bouwens \& Silk 1996; Driver \& Phillips 1996). A steep slope 
($\alpha<-1.8$) is only necessary 
to reconcile predicted number counts with observations in a flat Universe. 
A shallower slope is possible in other cosmologies. In the following, we adopt 
$\alpha_{\tau}=-1.3$ for bursting galaxies. Note that this translates into a steeper 
slope {\em of 
the LF} ($\alpha=-1.73$), as already noticed by Hogg \& Phinney (1997) 
for bursts.
The LF is normalized to agree with the UV counts and the
Cowie et al. (1996) redshift distribution.
Table~\ref{FL} gives the parameters of the SFR function and those 
of a Schechter fit to the 
corresponding luminosity function from $b_{\mathrm{j}}=-19$ to $-13$.
\begin{table}
\begin{tabular}{cccc}
\hline
Galaxy type & $M^{\ast}_{\rm b_j}$ & $\alpha$ & $\phi^{\ast}$ 
$(\mathrm{Mpc^{-3}})$\\
\hline
E & -20.02 & -1. & $1.91\,10^{-3}$\\ 
S0 & -20.02 & -1. & $1.91\,10^{-3}$\\
Sa & -19.62 & -1. & $2.18\,10^{-3}$\\
Sb & -19.62 & -1. & $2.18\,10^{-3}$\\
Sbc & -19.62 & -1. & $2.18\,10^{-3}$\\
Sc & -18.86 & -1. & $4.82\,10^{-3}$\\
Sdm & -18.86 & -1. & $9.65\,10^{-3}$\\
Bursting galaxies & -16.99 & -1.73 & $1.03\,10^{-2}$\\
\hline
& $\tau^{\ast}$ $(\mathrm{M_{\odot}\,Myr^{-1}})$ & $\alpha_{\tau}$ & $\phi^{\ast}_{\tau}$
$(\mathrm{Mpc^{-3}})$\\
\hline
Bursting galaxies & $3.95\,10^5$ & -1.3 & $6.63\,10^{-2}$\\
\hline
\end{tabular}
\caption{Luminosity functions parameters ($H_0=100\,\ksM$).
For bursting galaxies, we give both the SFR function and a Schechter fit to 
the LF from $b_{\mathrm{j}}=-19$ to $-13$.}
\label{FL}
\end{table}
\section{Galaxy counts}
Predictions of galaxy counts, with the bursting population ad\-ded to normal 
galaxies, have been computed in three 
cosmologies: an open Universe
($\Omega_0=0.1, \Lambda_0=0, H_0=65\,\ksM$), a flat Universe 
($\Omega_0=1, \Lambda_0=0, H_0=50\,\ksM$), and a flat, $\Lambda$-dominated,
Universe ($\Omega_0=0.3, \Lambda_0=0.7, H_0=70\,\ksM$).
The value of $H_0$ is chosen in each cosmology in order to obtain a 
13~Gyr-old Universe. 

The spectral evolution of normal galaxies is modelled as 
in FRV\footnote{A constant SFR is assumed for Sd-Im galaxies.}
and takes into account the stellar emission, the nebular emission 
and the extinction. A star formation scenario fitting the observed optical
to near-infrared spectral energy distribution of nearby templates 
is adopted for each Hubble type, with a 
star formation timescale increasing from the early to the late-types.
The age of each scenario determines the redshift of 
formation ($z_{\mathrm{for}}\sim 5-10$ for spheroidals 
and early-spirals, and $z_{\mathrm{for}}\sim 2$ for late-spirals).

\begin{figure*}
\resizebox{\hsize}{!}{\rotatebox{-90}{\includegraphics{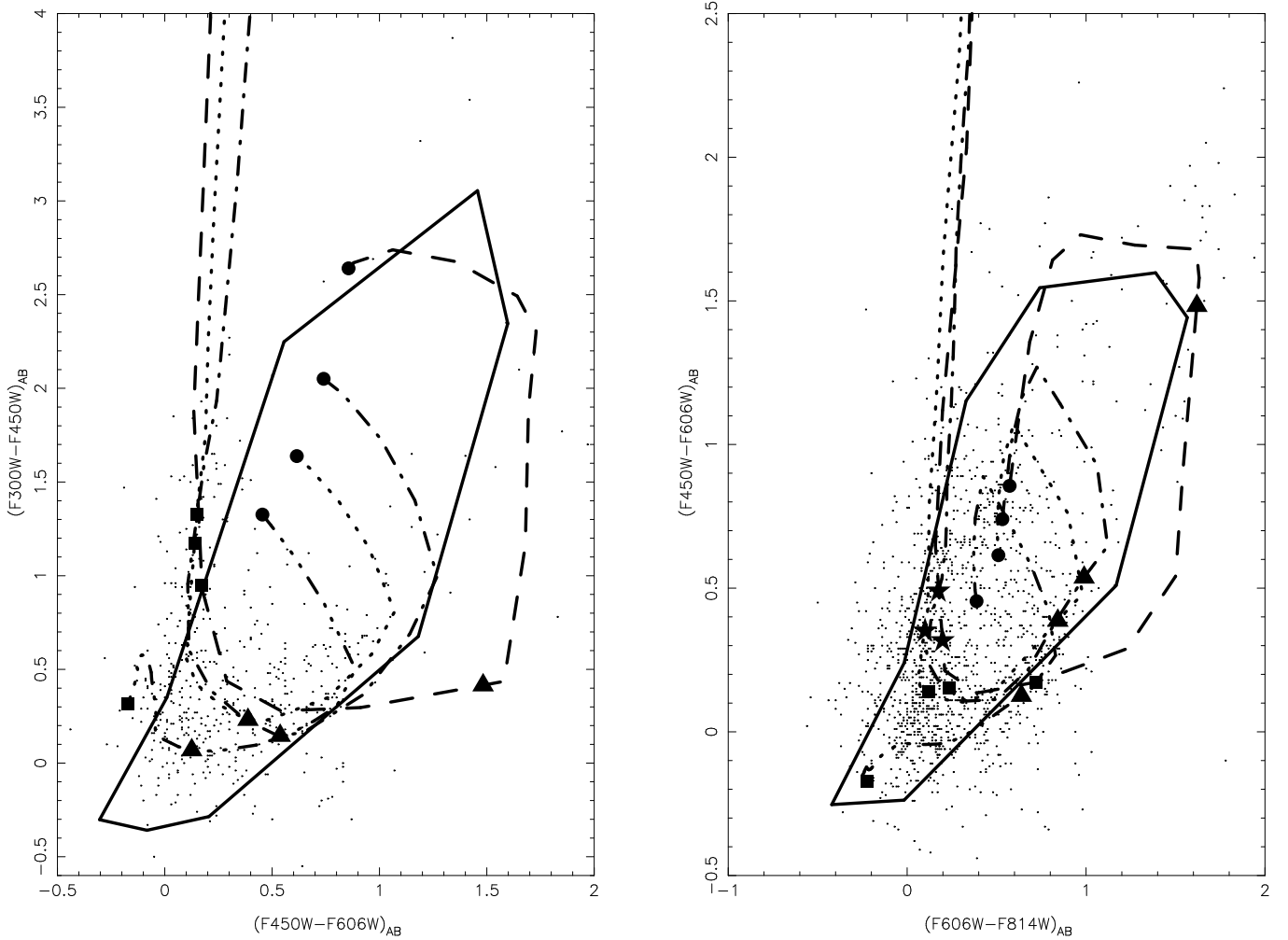}}}
\caption{Color-color plots of the galaxies in the HDF. 
The small dots are the observational data of Williams et al. (1996). The lines 
are the colors as a function of redshift of an elliptical (dashes),
an Sa (dash-dot), an Sbc (dots) and an Sd (dash-3 dots) in the case of an open
Universe. The filled circles, triangles, squares and stars correspond to 
the redshifts 0, 1, 2 and 3 respectively.
The solid polygon 
delimits the area covered by starbursting galaxies.}
\label{Lyman}
\end{figure*}

Especially important for predicting galaxy counts (e.g. 
Po\-zet\-ti et al. 1998) is the fact that {\em all} the
light shortward of the Lyman break is absorbed, either by the gas or by the
dust surrounding star-forming regions, in agreement with the observations 
of the Lyman continuum by Leitherer et al. (1995).
For high-$z$ galaxies, this feature is redshifted in the optical and produces
UV-dropouts (see Fig.~\ref{Lyman}) similar to those observed in the HDF (Steidel et 
al. 1996, Madau et al. 1996).
Although such UV-dropouts are also caused by the intergalactic medium
(Madau 1995), the predicted counts will essentially be the same.
We neglect the depression of the spectrum between 912~\AA\ and 1215~\AA\
due to the blanketing opacity of Lyman series lines. At $z<3$, the absorption
shortward of 912~\AA\ is by far more important, as evidenced 
by the figure~3 of Madau et al. (1996). Only $\sim$20~\% 
of the galaxies
are at higher redshift at $b_{\mathrm{j}}=28$, according to our  predictions,
and neglecting the additional absorption in the [912~\AA, 1215~\AA] 
wavelength range should not change significantly the predictions.

To ensure consistency with HCEB's LF, the LF of `normal' 
galaxies is constructed in the following way: bursting 
galaxies are classified into the three broad {\em spectral} types defined by 
HCEB, according
to their [O\,\textsc{ii}] equivalent width. For each class, the LF of `normal' 
galaxies is then added to that of bursting galaxies and its parameters
(assuming $\alpha=-1$) are fitted to the local step-wise LF of HCEB.
We prefer to use the step-wise LF, as the Schechter
parameterization of HCEB as a function of redshift poorly fits the data
near $M^{\ast}$ at $z\sim 0$.
Characteristics of the LF finally 
adopted are given in Table~\ref{FL}.
\begin{figure}	
\resizebox{\hsize}{!}{\rotatebox{-90}{\includegraphics{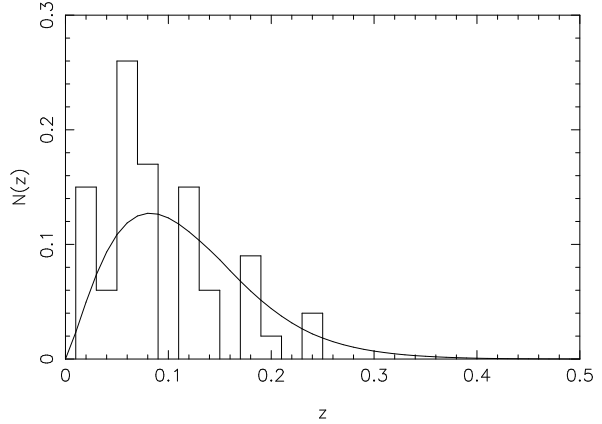}}}
\caption{Redshift distribution at $UV<18.5$. The histogram data are
from Treyer et al. (1998).}
\label{nzUV}
\end{figure}

Galaxy counts, color and redshift distributions at bright UV magnitudes are 
plotted on Figs.~\ref{UV} and~\ref{nzUV} and compared to the data of 
Armand \& Milliard (1994) and the preliminary results of Treyer et al. (1998). 
An open Universe is adopted for bright counts, which anyway depend 
weakly on cosmology. 
Though faint in the blue, bursting galaxies contribute strongly to the  UV 
bright counts thanks to their 
blue $UV-B$ colors, improving significantly the fits 
to the UV number-magnitude counts and color distributions (Fig.~\ref{UV}, solid lines). 
The agreement with the redshift distribution of Treyer et al. (1998) is
also reasonably good (Fig.~\ref{nzUV}).
\begin{figure*}
\resizebox{\hsize}{!}{\rotatebox{-90}{\includegraphics{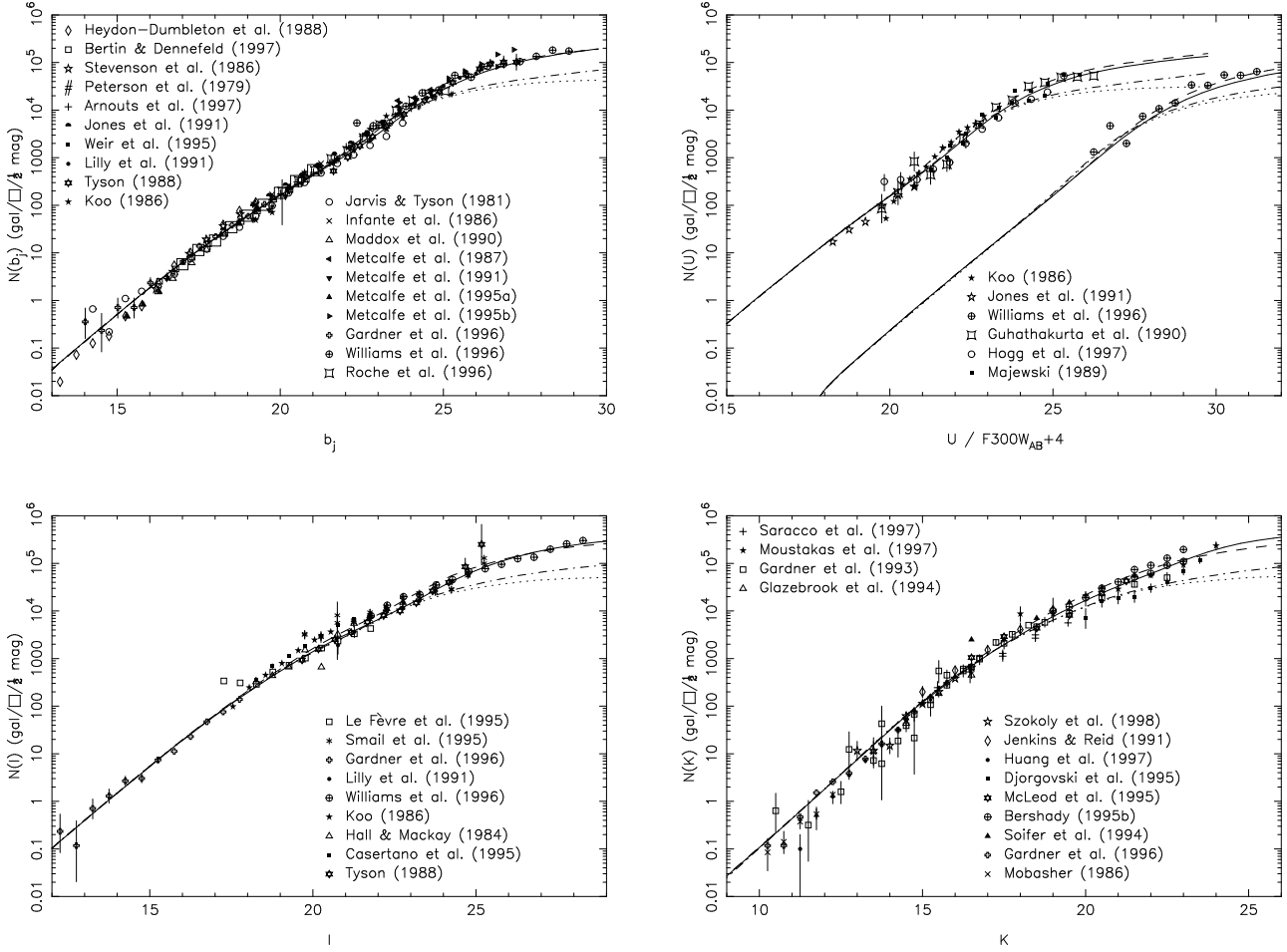}}}
\caption{Number magnitude counts in $b_{\mathrm{j}}$, $U$ (and $F300W$), $I$ 
and $K$. Predictions including bursting galaxies are plotted for an open 
Universe ($\Omega_0=0.1$, $\Lambda_0=0$, $H_0=65\,\ksM$: solid), a flat Universe
($\Omega_0=1$, $\Lambda_0=0$, $H_0=50\,\ksM$: dot-dashed) 
and a flat, $\Lambda$-dominated, 
Universe ($\Omega_0=0.3$, $\Lambda_0=0.7$, $H_0=70\,\ksM$: dashed). 
The predictions without bursting galaxies are also plotted as a dotted line
in the case of a flat, $\Lambda_0=0$, Universe.}
\label{Nm}
\end{figure*}

\begin{figure}
\resizebox{\hsize}{!}{{\includegraphics{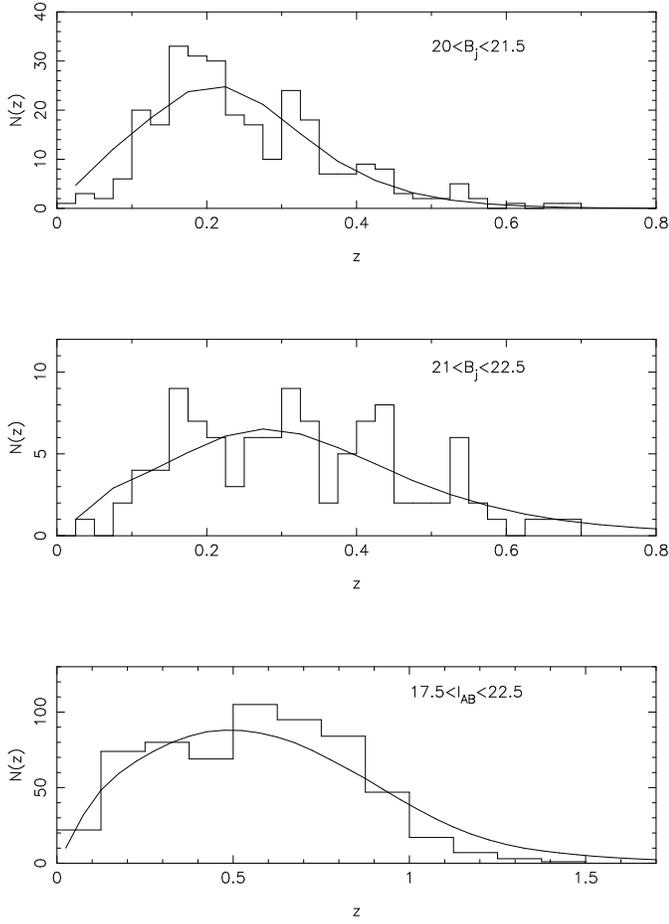}}}
\caption{Predicted redshift distributions (lines) in the case of an 
open Universe,
compared to the observations (histograms)
of Broadhurst et al. (1988; top), Colless et al. (1993; middle) and 
Crampton et al. (1995, Canada France Redshift Survey; bottom).}
\label{nzdivers}
\end{figure}

\begin{figure}
\resizebox{\hsize}{!}{\rotatebox{-90}{\includegraphics{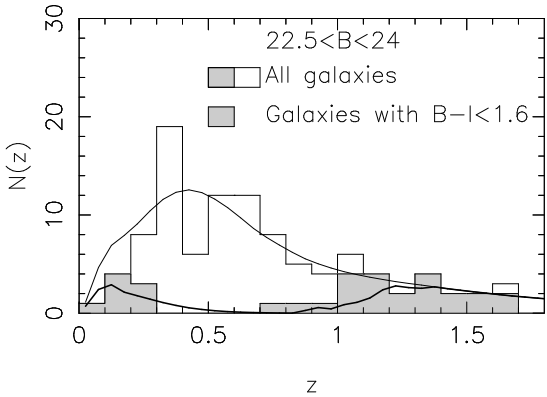}}}
\caption{Predicted redshift distribution ($22.5<B<24$) in the case of an 
open Universe, compared to the observations
of Cowie et al. (1996). The thick line is for galaxies with $B-I<1.6$ and
the thin line for all galaxies.}
\label{nz}
\end{figure}

\begin{figure}
\resizebox{\hsize}{!}{\rotatebox{-90}{\includegraphics{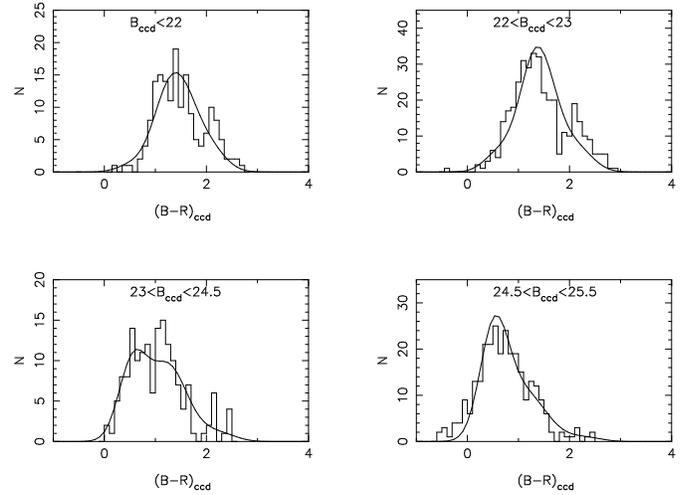}}}
\caption{Predicted color distributions (lines) in the case of an open Universe,
compared to the data (histograms) of Metcalfe et al. (1991, 1995a) in
several bins of magnitude.}
\label{nc}
\end{figure}

\begin{figure}
\resizebox{\hsize}{!}{{\includegraphics{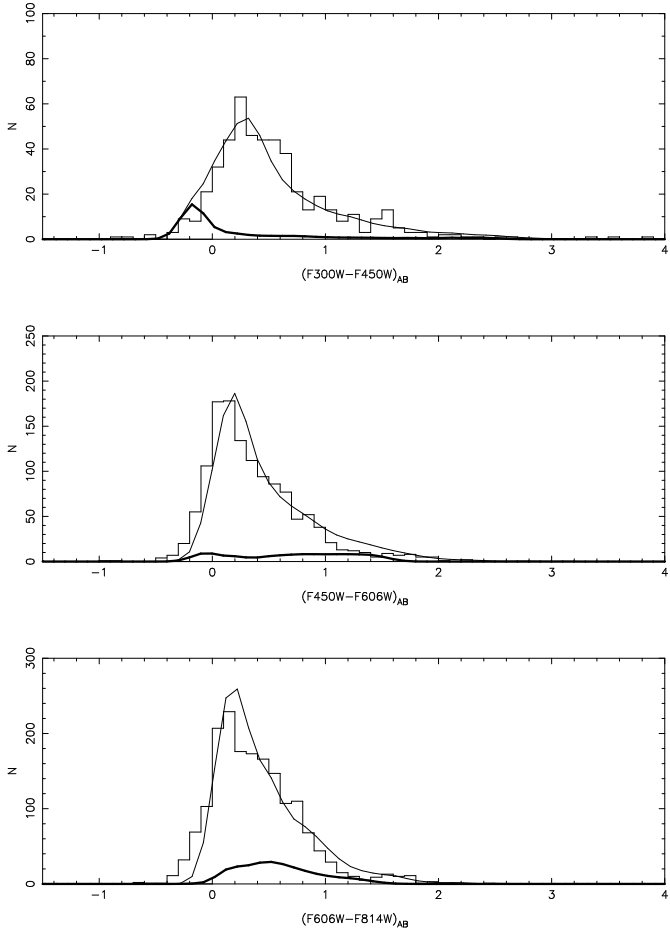}}}
\caption{Color distributions in the HDF for 
$(F300W)_{\rm AB}<27.75$, $(F450W)_{\rm AB}<28.75$, $(F606W)_{\rm AB}<29.25$
and  $(F814W)_{\rm AB}<28.75$ at 80 per cent 
completeness. 
The histograms are the data of Williams et al. (1996).
Predictions in the case of an open Universe are plotted for all 
galaxies (thin line) and for bursting galaxies only (thick line).}
\label{HDF}
\end{figure}

The geometry of the Universe becomes important at fainter magnitudes.
Galaxy counts in $b_{\mathrm{j}}$, $U$, $F300W$, $I$ and $K$, redshift and
color distributions
are plotted on Fig.~\ref{Nm}, \ref{nzdivers}, \ref{nz}, \ref{nc} and 
\ref{HDF}. 
The contribution of bursting galaxies to counts at longer wavelengths is 
much smaller than in the UV. They represent less than 10 per cent of the total 
number of galaxies at $B=22.5-24$ 
in the Cowie et al. (1996) redshift survey and cannot be the main explanation
of the excess of faint blue galaxies observed over the no-evolution predictions.
High-redshift,
intrinsically bright galaxies forming stars at a higher rate in the past are 
the most likely explanation. They correspond to the $z>1$ tail of the Cowie et al. (1996)
redshift distribution and are well modelled with PLE scenarios (Fig.~\ref{nz}).
In an open or a flat, $\Lambda$-dominated, Universe, PLE scenarios
reproduce the $b_{\mathrm{j}}$, $U$, $I$ and $K$ counts (Fig.~\ref{Nm}), assuming a normalization
of the LF to the bright counts of Gardner (1996) as discussed 
in FRV. They also fit the redshift and color distributions 
(Figs.~\ref{nzdivers} to \ref{HDF}).

The agreement with the blue counts in the Hubble Deep Field (Williams et al. 1996) is notably 
satisfying. Though a small deficit may be observed in the $F300W$ band 
(3000 \AA),
the $F300W-F450W$ (3000~\AA--4500~\AA) color distribution is well reproduced 
(Fig.~\ref{HDF}). 
The fraction of vBG at these faint magnitudes
is small; they are therefore not the main reason for the agreement 
with HDF data.
Since the volume element at high redshift is lower in a flat, $\Lambda=0$, Universe,
the contribution of bursting galaxies relative to normal galaxies is higher,
but, {\em as modelled here}, they
are unable to reproduce the faint counts in any of the four bands
(dotted and dot-dashed lines, Fig.~\ref{Nm}).
\section{The angular correlation function}
\begin{figure}
\resizebox{\hsize}{!}{\rotatebox{-90}{\includegraphics{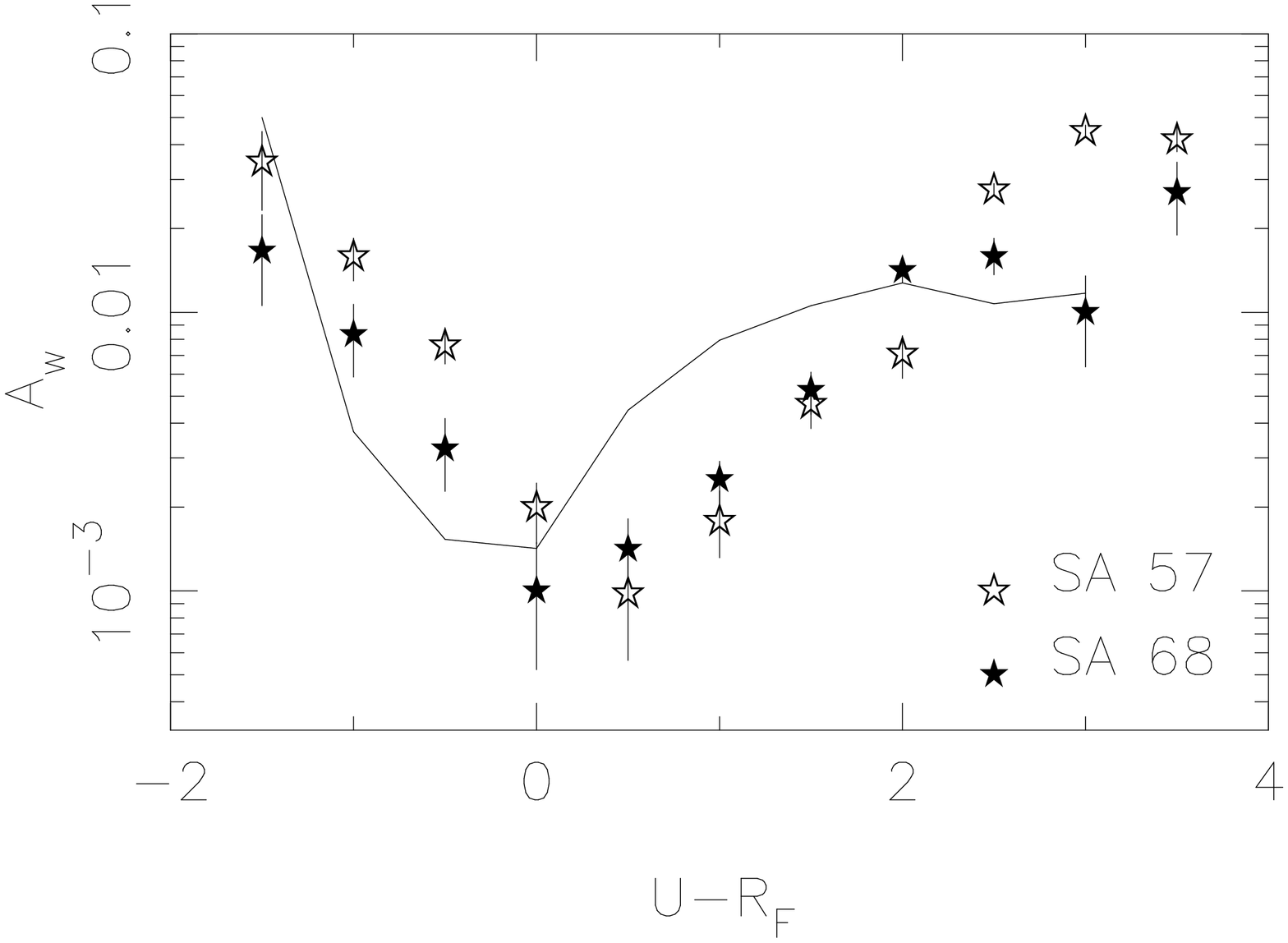}}}
\caption{Amplitude at $1\degr$ of the angular correlation function in $B_{\rm J}=20-23.5$,
as a function of $U-R_{\rm F}$ color in bins of 1 magnitude wide. Stars are 
from Landy et al. (1996). The solid line is the amplitude predicted
without evolution of the intrinsic clustering in proper coordinates.
%The two-point correlation function is $\xi(r)(r/r_0)^{-\gamma}$
}
\label{Aw}
\end{figure}

The angular correlation function might be a useful constraint on bursting 
galaxies, as it is directly related 
to the redshift distribution. In a $B_{\rm J}=20-23.5$ sample, 
Landy et al. (1996) recently obtained an 
unexpected increase of the amplitude ($A_{\mathrm{w}}$)
of the angular correlation function with galaxy colors 
$U-R_{\rm F}<-0.5$, and suggested that this might be due to a population of 
vBG located at $z<0.4$. 
We have computed $A_{\mathrm{w}}$ from our redshift distributions, assuming a classical
power law ($\xi(r,z)=(r_0/r)^{\gamma}(1+z)^{-(3+\epsilon)}$)
for the local spatial correlation function and no evolution of the intrinsic 
clustering in proper coordinates ($\epsilon=0$).
A slope $\gamma=1.8$ and a single correlation length 
$r_0=5.4h^{-1}\,{\rm Mpc}$ (see Peebles 1993) have been adopted for all types. 
The increased correlation in the blue naturally arises from our computations 
(Fig.~\ref{Aw}) and is due to the population of bursting galaxies in our model.
The interval of magnitudes, the faint $M^{\ast}$, 
and the color criterion conspire to select
galaxies in a small range of redshift.
In spite of the simplicity of our computation of $A_{\mathrm{w}}$, 
the trend we obtain is very 
satisfying. 
Increasing the complexity of the model might afford a better fit to the 
$A_{\mathrm{w}}$-color relation, but at the expense of a higher number of parameters. 
\section{Conclusion}
We have modelled a vBG population prominent in the bright UV counts
by assuming that they are galaxies that undergo cycles of star-formation activity.
A cycling star formation rate leads to very blue colors
in a more physical way than by assuming a population of unevolving galaxies.
Our modelling fits the 2000~\AA\ bright counts (Armand \& Milliard 1994), 
the redshift survey of Cowie
et al. (1996) and the angular correlation function of Landy et al. (1996). 
The nature of vBG is poorly constrained from the data studied here, 
but we tentatively identify them 
from their typical luminosities and ${\rm H}\alpha$ equivalent widths 
($\sim 200$ \AA) with H\,\textsc{ii} galaxies (Coziol 1996). Such a conclusion 
is reinforced by observations of vBG  at $z\sim0.1-0.35$
with low luminosities, but high surface 
brightnesses, and intense emission lines (Koo et al. 1994).
Multispectral observations of individual galaxies, from the UV to the 
infrared (e.g. Almoznino \& Brosch 1998), 
should help to determine the possible contribution of the 
underlying population of old stars, and then the star formation history.

Very blue galaxies, as modelled in this paper, are only a small 
fraction of the number of
galaxies predicted at faint magnitudes in the visible and are not 
the main reason
for the excess of faint blue galaxies, although they may cause some confusion
in the interpretation of the faint surveys. In an open
or a flat, $\Lambda$-dominated, Universe, the population
of normal high redshift star-forming galaxies, 
even with a nearly flat LF, reproduces fairly well the counts down to the faintest magnitudes
observed by the Hubble Space Telescope. 

As is now well established, this population is unable to 
explain the excess of faint blue galaxies in a flat Universe (e.g. Metcalfe et
al. 1996). 
Recent morphological surveys have revealed a higher number of irregular/peculiar
galaxies (e.g. Gla\-ze\-brook et al. 1995; Odewahn et al. 1996) with smaller sizes at faint magnitudes
than expected from PLE models. Although these results depend sensitively 
on the evolution of the apparent UV morphology
(Abraham et al. 1996), which may make galaxies 
look more patchy (O'Connell 1997) and even split them in several pieces at high redshift (Colley et al. 1996), 
they seem to favor
bursting dwarfs as a possible explanation of the blue excess in a flat 
Universe. Simply changing the local luminosity function may, however,
not be the solution, since a significant steepening of the slope would lead 
to an excess of galaxies at very low redshift, whereas a much higher 
normalization would be in disagreement with UV counts. 
An alternative is to produce stronger bursts at high redshift. This would
brighten them, but would also raise the number
of old stars in them and make it more difficult to obtain very blue colors
at low redshift. If the blue excess is really due to bursting galaxies, 
the hypothesis of the conservation of the number of galaxies must be relaxed.
Either new galaxies are continually formed or galaxies have merged.
In the first case, many old red remnants must have been produced and
might be detected by future near-infrared surveys, while in the second 
case, many merging galaxies should be observed in the far-infrared 
by the Infrared Space Observatory.
\begin{acknowledgements}
We are particularly grateful to Malcolm  Bremer for carefully reading
the manuscript. We   acknowledge  fruitful discussions  with Bruno Milliard 
and Jos\'e Donas and we thank them for providing us with details on the FOCA
experiment. 
We also thank Joe Silk and Gus
Evrard  for    their  remarks   and  recommendations    regarding  the
continuation of this work. M. F. acknowledges partial support from the
National Research Council through the Resident Research Associateship Program.
\end{acknowledgements}

\end{document}